\documentclass[wcp]{jmlr} 

\usepackage{booktabs}
\usepackage[load-configurations=version-1]{siunitx}
\usepackage{microtype}
\usepackage{multirow}
\usepackage{tikz}
\usetikzlibrary{automata,positioning}
\usepackage{amsmath}
\usepackage[super]{nth}
\usepackage{kbordermatrix}

\title{Learning Several Languages from Labeled Strings\titletag{: \\ State Merging and Evolutionary Approaches}}

\author{\Name{Alexis Linard} \Email{a.linard@cs.ru.nl} \\
       \addr Institute for Computing and Information Science\\ Radboud University \\ Nijmegen, The Netherlands
       }

\jmlrpages{}
\jmlrproceedings{}{}

\begin{document}

\maketitle

\vspace*{-0.5cm}

\begin{abstract}
The problem of learning pairwise disjoint deterministic finite automata (DFA) from positive examples has been recently addressed.
In this paper, we address the problem of identifying a set of DFAs from labeled strings and come up with two methods.
The first is based on state merging and a heuristic related to the size of each state merging iteration.
State merging operations involving a large number of states are extracted, to provide sub-DFAs.
The second method is based on a multi-objective evolutionary algorithm whose fitness function takes into account the accuracy of the DFA w.r.t. the learning sample, as well as the desired number of DFAs. We evaluate our methods on a dataset originated from industry.
\end{abstract}


\section{Introduction}
\label{sec:intro}
The problem of identifying a set of DFAs from examples that belong to different unknown languages is a recent task in grammatical inference, explored in \cite{learnaut} for the first time. The setting considered was the observation of positive examples from multiple disjoint regular languages. Considering, for example, the following set of strings $S = \{aa,aaa,aaaa,abab,ababab,abba,abbba,abbbba\}$, the claim is that gathering a single automaton might be less informative than gathering several DFAs respectively encoding the languages $a^*$, $(ab)^*$ and $ab^*a$. There is a trade-off between the number of languages and how specific each language should be. That is, covering all words through a single language may not be the desired result, but having a language for each word may also not be desired. The approaches explored so far are based on clustering, either using clustering by compression as pre-processing, then applying DFA inference algorithms; or identifying tandem repeats in the strings and then generalizing the found patterns.

In this paper, we consider a setting where one can observe labeled (i.e. positive and negative) examples from multiple languages. Hence, classical DFA inference algorithms, such as Regular Positive Negative Inference (RPNI) in \cite{oncina1992identifying} or Evidence-Driven State Merging (EDSM) in \cite{lang1998results} can be extended. These algorithms consist of computing a Prefix-Tree Acceptor (PTA) for all the input strings and marking the states as rejecting or accepting depending on the input words. Then, state merging is iteratively applied, to finally obtain an automaton consistent with the dataset. Many modifications of these seminal state merging algorithms have been explored, such as adding merge constraints to state merging \citep{lambeau2008state}. Extensions to Probabilistic Deterministic Finite Automata \citep{carrasco1994learning}, and grammatical inference through SAT \citep{heule2010exact} or SMT solvers \citep{Neider2012,smetsers2017grammatical} have been explored. Nonetheless, none of these state merging algorithms have been modified to output more than one DFA, as in the spirit of \cite{learnaut}.

We propose here a modification of state merging algorithms, in particular RPNI, which outputs a set of DFAs.
We also propose an evolutionary algorithm inspired by \cite{dupont94} whose fitness function takes into account the accuracy of the DFA w.r.t. the learning sample -- that is to say ensuring that all positive strings are accepted and all negative strings rejected -- while in addition taking into account the desired number of DFAs. 

This paper is organized as follows. In Section \ref{sec:def} we recall preliminary definitions.
We present then, in Sections \ref{sec:ptasplit} and \ref{sec:evolution} our techniques to infer several languages from a set of labeled strings.
In Section \ref{sec:experiments} we show the results we achieved on an industrial case study. 
Finally, we discuss and conclude about the proposed methods.

\section{Definitions}
\label{sec:def}

In this section, we present the different concepts and definitions related to the learning of a set of DFAs from labeled strings.

\paragraph{Strings}
Let $\Sigma$ denote a finite alphabet of \emph{symbols}. 
A \emph{string} $x = a_0 \ldots a_n$ is a finite sequence of symbols.
The empty string is denoted $\epsilon$.
We denote by $\Sigma^{\ast}$ the set of all strings over $\Sigma$, and by $\Sigma^{+}$ all nonempty strings over $\Sigma$ (i.e.\ $\Sigma^{\ast} = \Sigma^{+} \cup \{ \epsilon \}$).
Similarly, we denote by $\Sigma^i$ the set of all strings over $\Sigma$ of length $i$.
Given a set of strings $S$, we denote by $\texttt{prefix}(S)$ the set of prefixes of strings in $S$.

\paragraph{Labeled examples}
A set of \emph{labeled strings} is a finite set of strings divided into positive strings (belonging to the language to learn) and negative strings (not belonging to the language to learn). We refer to $S_+$ and $S_-$ as the set of positive and negative examples.

\paragraph{Languages}
A regular language can be described by a \emph{deterministic finite automaton} (DFA), which is a tuple $\mathcal{A} = (\Sigma, Q, q_0, \delta, F)$, where $Q$ is a finite set of \emph{states}, $q_0 \in Q$ is the \emph{starting state}, $\delta: Q \times \Sigma \to Q$ is a partial \emph{transition function} from states and symbols to states, and $F \subseteq Q$ is a set of \emph{accepting states}.
Note that the transition function can also be represented as a \emph{transition matrix} $M$, being a two-dimensional array of size $|Q|\times|\Sigma|$ where $M_{i,j} = \delta(q_i,a_j)$ with $q_i$ and $M_{i,j} \in Q$ and $a_j \in \Sigma$; and the set of accepting states as an \emph{output array} $O$ of size $|Q|$ where the $i$-th value of the array would be 1 if $q_i$ is accepting, 0 otherwise.
Let $q$ be a state, $a$ be a symbol and $u$ be a string. Then we extend $\delta$ to $\delta^{\ast}: Q \times \Sigma^{\ast} \to Q$ by $\delta^{\ast}(q, \epsilon) = q$ and $\delta^{\ast}(q, au) = \delta^{\ast}(\delta(q, a), u)$.
The set of tails described by a state $q$ is the set $\{ x \in \Sigma^* : \delta^{\ast}(q, x) \in F\}$.
Hence, the language $\mathcal{L}(\mathcal{A})$ recognized by DFA $\mathcal{A}$ is the set of tails of the starting state of $\mathcal{A}$.
A Prefix Tree Acceptor (PTA) of an non empty set $S_+$ is a tree-like DFA $\mathcal{A} = (\Sigma, Q, \epsilon, \delta, S_+)$ such that $Q = \texttt{prefix}(S_+)$ and $\forall ua \in \texttt{prefix}(S_+) \ :\ \delta(u,a) = ua$.

\section{State-merging technique}
\label{sec:ptasplit}

Previous work \citep{learnaut} highlighted the interest of gathering more than one DFA as result of learning algorithms. In the setting we consider, we deal with both positive and negative examples. Hence our focus on a modification of RPNI.

\subsection{RPNI}

In RPNI, a PTA is built from the learning sample.
Then, the task is to try to generalize this prefix tree by merging states, always ensuring that the resulting automaton is consistent with both negative and positive samples. 
At any moment of the iterative process of state merging, states are categorized, i.e. marked as either \textit{red} (states that have been analyzed/merged, and that will be part of the final automaton) or \textit{blue} (candidate states, that need to be analyzed to see if merging with a \textit{red} state is possible).
RPNI can be seen as based on this \textit{``red/blue''} framework \citep{cdlh}, which is an iterative process with the following operations: \textit{compatibility}: operation ensuring that a blue and a red state are either both accepting or rejecting; \textit{merging}: operation iteratively performed, consisting of merging one blue and one red state that are compatible. Merging is possible iff the resulting DFA is consistent with $S_+$ and $S_-$; \textit{promoting}: operation turning a \textit{blue} state into \textit{red} in case the \textit{blue} state has been fully analyzed, and it cannot be merged with any of the \textit{red} states; and \textit{folding}: operation enabling to reconnect a possibly disconnected part of the DFA after merging two states


\subsection{RPNI-Splitting}

While performing state merging, it appears that some merges are involving either a considerable number of states (relative to the overall size of the DFA at the current iteration), or a considerable number of accepting states (in comparison with the size of $S_+$).
The approach assumes the following: in case a merging operation involves \textit{many} states, that is to say, that the size of the DFA drastically changes, then the part being merged has a remarkably distinct meaning with regards to the rest of the DFA. That is, in case a significant number of accepting states collapse all at once, it might be interesting to look into the related strings belonging to $S_+$ by deriving a separate solution from this sub-sample. We say that we \textit{split} the DFA at this iteration.
The notion of \textit{big merge} can be defined as a function of an input \textit{splitting} parameter $k$, the larger the more sub-DFAs would be returned.


In Algorithm \ref{algo:rpnisplit}, we present a modification of RPNI to output a set of DFAs.
The definitions for $\texttt{buildPTA}$, $\texttt{rpniCompatible}$, $\texttt{merge}$ and $\texttt{promote}$ are extensively described in the literature (see for instance \cite{cdlh}). The function $\texttt{choose}$ returns the first blue state in alphabetical order.
We refer to RPNI without splitting as $\texttt{standardRPNI}$.
$|Q_{m}|$ stands for the number of accepting states being merged during an iteration of state merging, $|S_+|$ the size of the positive sample, $|Q|-|Q_{m}|$ for the difference of size of the DFA before and after an iteration of state merging.
As soon as a consequent merge is detected, we extract all the strings $s \in F \backslash F_m$ to derive a fine-grained subsidiary solution. The idea is to build a sub-PTA accepting all strings $s \in F \backslash F_m$ and rejecting $S_-$. Note here all the DFAs reject $S_-$. Then, 1. usual state merging is executed, to gather a proper sub-DFA from the particular set of strings originated from the \textit{big merge} extraction and 2. RPNI-splitting is recursively called with $S_+ \backslash F_m$ as positive sample and $S_- \cup F_m$ as negative sample, which means using the strings in $S_+$ which have not been merged as positive sample, and adding the merged ones to the negative dataset.
At the end of the overall process, we gather as many sub-DFAs as detected \textit{big merges}. The main advantage of the application of this heuristic is its genericity: it is applicable to other state merging algorithms such as EDSM.
We present an example of \texttt{RPNI-Splitting} in Appendix \ref{app:ex-rpni}.

\begin{algorithm2e}[!h]
\caption{RPNI-splitting.}
\label{algo:rpnisplit}
 \DontPrintSemicolon
\KwIn{$S_+,\ S_-$: set of labeled strings, $k$: splitting parameter}
\KwOut{set of at most $k$ DFAs accepting all strings $s \in S_+$ and rejecting all strings $r \in S_-$}
$\mathcal{A} \gets \texttt{buildPTA}(S_+)$\;
$\mathcal{R} \gets \{q_{\epsilon}\}$, $\mathcal{B} \gets \{q_{a} : a \in \Sigma \cap \texttt{prefix}(S_+)\}$  \tcp*{set of red/blue states}  
\While{$\mathcal{B} \neq \emptyset$}{
	$\texttt{choose}(q_b \in \mathcal{B})$\;
    $\mathcal{B} \gets \mathcal{B} \backslash \{q_b\}$\;
    \If{$\exists q_r \in \mathcal{R}$ \textbf{s.t.} $\texttt{rpniCompatible}(\texttt{merge}(\mathcal{A},q_r,q_b),S_-)$}{
    	$\mathcal{A}_{m} \gets \texttt{merge}(\mathcal{A},q_r,q_b)$\;
    	\If{$k>1$ \textbf{and} ($|F \backslash F_m| \geq \frac{|S_+|}{k}$ \textbf{or} $|Q|-|Q_{m}| \geq \frac{|Q|}{k}$)}{
            \textbf{return} $\{ \texttt{standardRPNI}(F \backslash F_m, S_-) \} \cup \texttt{RPNI-splitting}(S_+ \backslash F_m, S_-\cup F_m, \frac{k}{2})$\;
        }
        $\mathcal{A} \gets \mathcal{A}_{m}$\;
        $\mathcal{B} \gets \mathcal{B} \cup \{\delta(q,a) \in Q \backslash \mathcal{R} : q \in \mathcal{R} \land a \in \Sigma\} $\;
    }
    \Else{
    	$\mathcal{R},\ \mathcal{B} \gets \texttt{promote}(q_b,\mathcal{R},\mathcal{B},\mathcal{A})$\;
    }
}
\textbf{return} $\{ \mathcal{A} \}$ \;
\end{algorithm2e}

\vspace*{-1.35cm}
\section{Evolutionary algorithm}
\label{sec:evolution}
\vspace*{-0.2cm}

Evolutionary algorithms (EAs) for learning DFAs are an interesting solution for the problem of grammatical inference. By construction, they ensure that the returned DFA is consistent with the input dataset since the proportion of strings correctly classified is part of the fitness function, with the \textit{best possible fitness value is attained} as stopping criterion. We propose here to base the selection not only on the accuracy of the DFA, but also on the number of desired DFAs $k$. Hence the multi-objectiveness of our algorithm.

\vspace*{-0.2cm}
\subsection{DFA learning through Evolution}
\vspace*{-0.1cm}

Evolutionary Algorithms for DFA learning have been investigated in \cite{dupont94} and consist of the following steps: 1. the generation of an initial population containing a single randomly generated DFA; 2. the evaluation of the fitness of each DFA in the population. Often, the accuracy of the DFA w.r.t. $S_+$ and $S_-$ is a measure of choice for fitness; 3. performing genetic operations on the individuals; 4. the selection of the best individuals in the population. Steps 2, 3 and 4 are repeated until at least one individual in the population achieves perfect fitness or a maximum number of allowed iterations (stopping criterion) is reached: then, the best DFA in the population is returned. 

Whenever more than one objective has to be met (accuracy, size of the DFA), multi-objective evolutionary algorithms such as  NSGA-II \citep{nsga2} can be used and consist in taking into account the fitnesses of the different objectives, then making a trade-off of these in the selection of the best individuals. Multi-objective optimization for DFA learning has already been used by \cite{rooijen} in an active learning setting.

\subsection{Evolutionary algorithm for multiple DFA learning}

\paragraph{Initial Population}
\label{subsec:initpop}
The initial operation is to define the seed population. We chose to start with an initial population consisting of one DFA $\mathcal{A}$ per string in $S_+$ such that $\forall s \in S_+,\ \mathcal{A} = \texttt{buildPTA}(s)$.

\paragraph{Genetic Operators} There are two main genetic operators that, given parent(s), generate new individuals using the parent(s) genome as a basis. Mutation consists of, given a parent DFA $p_1$, randomly mutating an entry in the transition matrix $M_{p_1}$ to a random value in $Q$. The same operation can be applied to the output array $O$. Crossover consists of taking 2 parents $p_1$ and $p_2$, mixing their genomes, and producing children solutions out of them. Concretely, a random index (single-point) is chosen in the transition matrices $M_{p_1},M_{p_2}$ of the parents from which the two sub-parts will either go to one child or to the other.

\paragraph{Fitness Functions}
\label{subsec:fitness}

After a new generation is created, two \textit{scores} are assigned to each individual: $f_1$ based on the accuracy $a$ of the DFA (proportion of strings in $S_+$ and $S_-$ which are correctly classified) and $f_2$ based on the number of sub-DFAs $n$ (see paragraph on transitions clustering) that are calculated out of the individual and given $S_+$.
\vspace*{-0.1cm}
\[
f_1 = 1-a \ \ \ \ \ \ \ \ \ f_2 = |1- \frac{n}{k}|
\]
\vspace*{-0.3cm}

Note here that the optimal fitnesses we want to reach are 0, so that the evolutionary algorithm will tend to discard individuals having the greatest scores.

\vspace*{-0.2cm}
\begin{algorithm2e}[!h]
\caption{Getting sub-DFAs out of a DFA by transition clustering.}
\label{algo:transitions}
 \DontPrintSemicolon
\KwIn{$S_+$ : set of positive strings, $\mathcal{A} = (\Sigma, Q, q_0, \delta, F)$ : input DFA}
\KwOut{$\mathcal{O}$ : set of sub-DFAs of $\mathcal{A}$}
$\mathcal{O} \gets \emptyset$\;
\For{$s \in S_+$}{
	\If{$\delta^*(q_0,s) \in F$}{
    	$Q_{sub} \gets \{\delta^*(q_0,p) : p \in \texttt{prefix}(s)\} $ \;
        $\forall q \in Q,\ \forall a \in \Sigma : \delta_{sub}(q,a) \gets \{\emptyset\}$  \tcp*{initialize $\delta_{sub}$}   
        $q_{curr} \gets q_0$ \;
        \For{$ua \in \texttt{prefix}(s)$}{
            $\delta_{sub}(q_{curr},a) \gets \delta^*(q_0,ua) $ \;
            $q_{curr} \gets \delta^*(q_0,ua)$ \;
        }
    	 $\mathcal{O} \gets \mathcal{O}\ \cup\ (\Sigma, Q_{sub}, q_0, \delta_{sub}, \delta^*(q_0,s))$\;
    }
}
\textbf{return} $\mathcal{O}$\;
\end{algorithm2e}
\vspace*{-0.4cm}

\paragraph{Transitions Clustering}
\label{subsec:evolution}

The number of sub-DFAs that can be rendered out of a single DFA is calculated by looking at the number of different paths taken by the strings in $S_+$ when parsing them in the DFA (Algorithm \ref{algo:transitions}).
Our goal is then to cluster the transitions as in the spirit of \cite{ZHANG20172353}.
In their work, they perform state sequence clustering to figure out different driving behaviors, represented as timed automata. 
That is why, given: 1. a DFA and 2. a positive dataset $S_+$, we first look which transitions are used by the strings in $S_+$, and then return the number of different \textit{paths} taken by the strings. Note here that the accepting state of the string is part of the solution.
We can then, at the termination of the evolutionary algorithm once the best individual is found, return a solution containing $k$ sub-DFAs using the $k$ set of transitions, as explained in Algorithm \ref{algo:transitions}. An example of transitions clustering is shown in Appendix \ref{app:transition-clust}.

Note here that a candidate DFA accepting none of the strings in $S_+$ contains no set of transitions from its initial state to an accepting state. As a consequence, the result returned by Algorithm \ref{algo:transitions} will be the empty set. From a practical perspective, the way to use the evolutionary algorithm is to apply Algorithm \ref{algo:transitions} to the best individual returned by the EA, to gather $k$ DFAs accepting strings in $S_+$ and rejecting all strings in $S_-$.

\vspace*{-0.3cm}

\section{Experiments}
\label{sec:experiments}

\vspace*{-0.2cm}

We evaluated our methods on a dataset inspired by the industrial case study described in \cite{learnaut}.
We consider here a set of 6 languages  $\mathcal{I} = \{\mathcal{L}(a^+), \mathcal{L}((ab)^{\geq 2}),$  $ \mathcal{L}((abc)^+),\mathcal{L}(ab^+a), \mathcal{L}(a^+b^+), \mathcal{L}(a(bc)^+a)\}$ from which 100 strings are generated. 
The strings are divided into a training set and a test set $T$.
The proportion of strings in the training set in comparison to the strings in the test set (density) varies from 0.02 to 0.20.
We consider 4 methods: the 2 methods presented in this paper, RPNI-splitting (RP) and the evolutionary algorithm (EA), as well as the 2 methods of \cite{learnaut}, clustering by compression (CC) and the identification of tandem repeats (TR).
We denote by $\mathcal{O}$ the set of DFAs returned by the methods.
The results are stated in terms of purity. This represents to what extent the learned DFAs recognize a single target language. Note here that this measure is relevant in the case of disjoint languages. For each learned DFA, we count the number of strings issued from the most common of input the languages, such that
\vspace*{-0.2cm}
\[
purity(T,\mathcal{O}) = \frac{1}{|T|} \sum_{i\ \in\ \mathcal{I}}\ \max_{o\ \in\ \mathcal{O}}\ |T \cap i \cap \mathcal{L}(o)|
\]

\vspace*{-0.5cm}
\begin{figure}[!h]

\subfigure[$k=2$]{
	\label{fig:k2}
    \hspace*{-0.3cm}
    \includegraphics[width=0.255\linewidth]{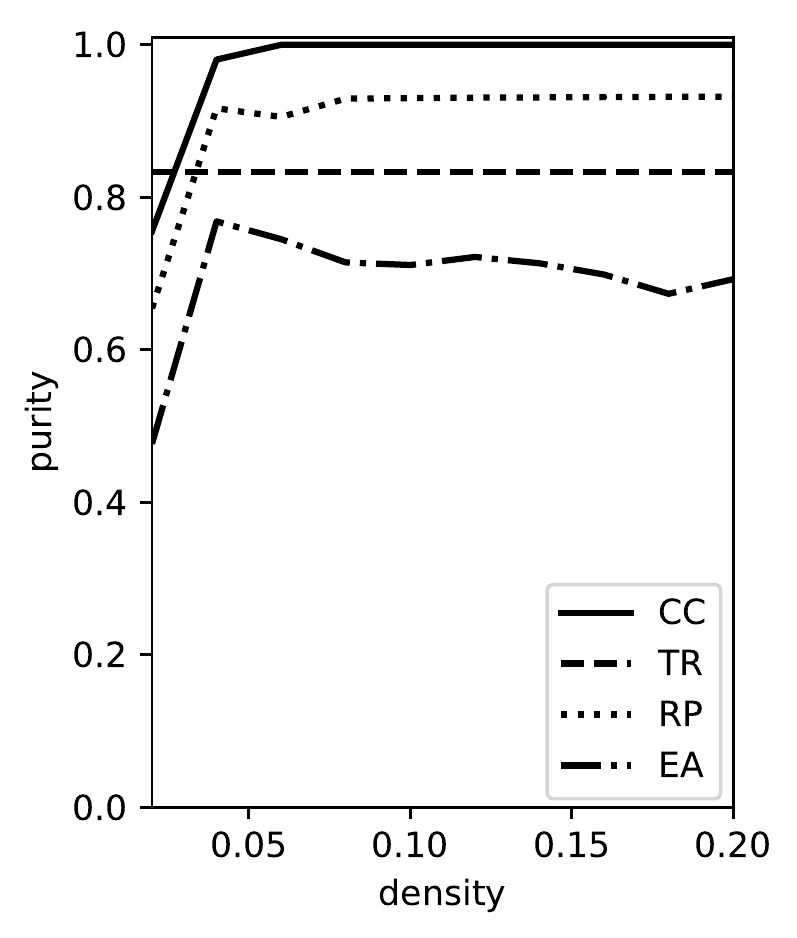}
    }
\subfigure[$k=3$]{
	\label{fig:k3}
    \hspace*{-0.5cm}
    \includegraphics[width=0.24\linewidth]{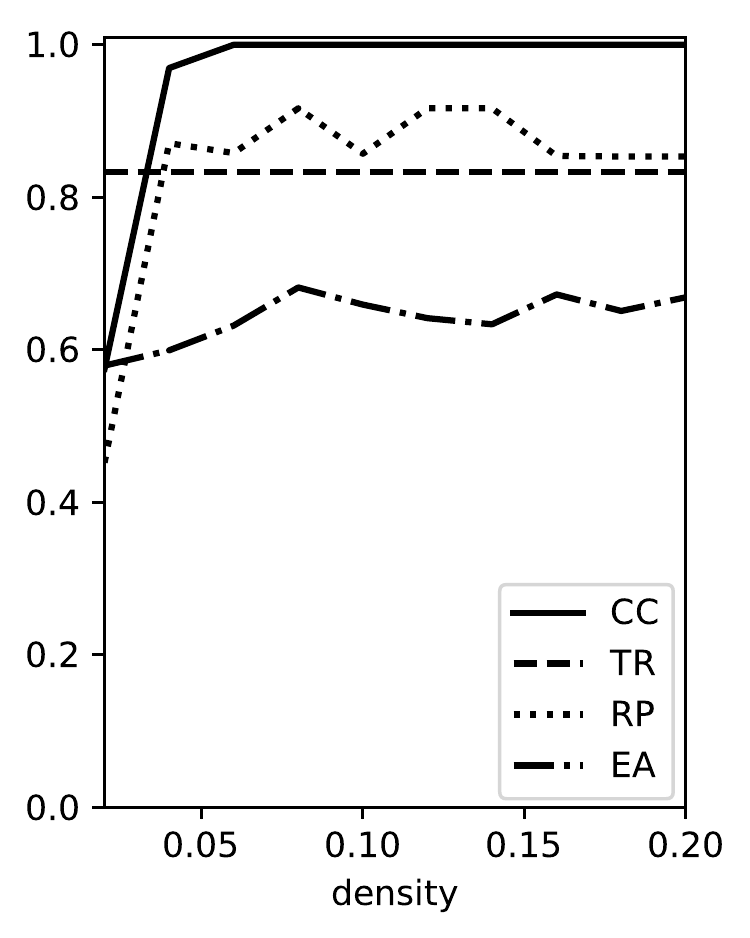}
    }
\subfigure[$k=4$]{
	\label{fig:k4}
    \hspace*{-0.5cm}
    \includegraphics[width=0.24\linewidth]{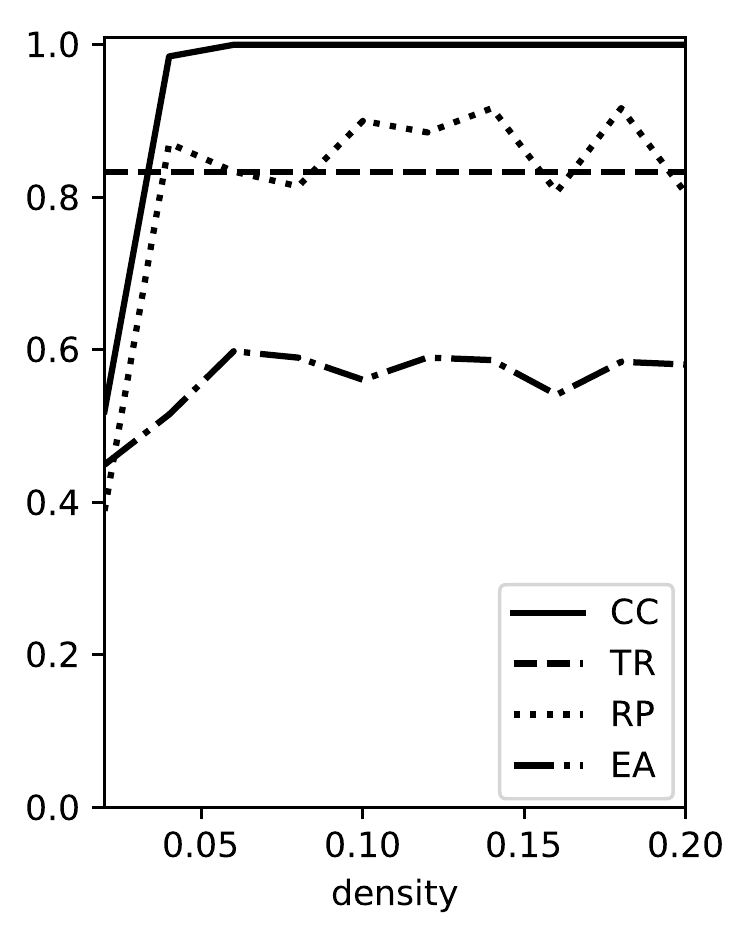}
    }
\subfigure[$k=5$]{
	\label{fig:k5}
    \hspace*{-0.5cm}
    \includegraphics[width=0.24\linewidth]{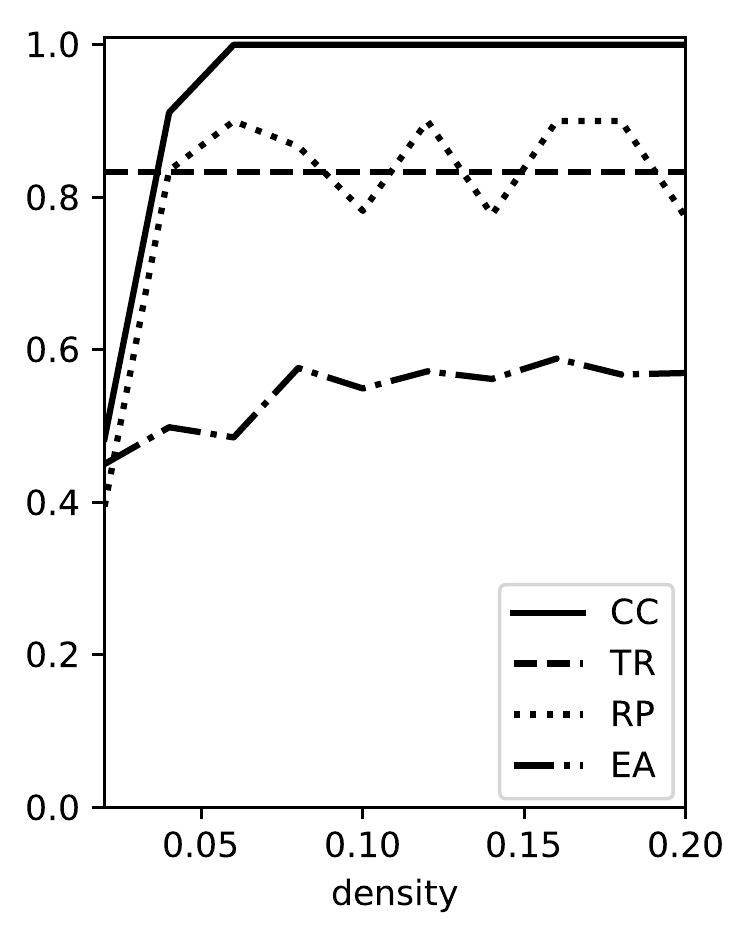}
    }
\vspace*{-0.3cm}
\caption{Results depending on the number of DFAs to learn.}
\label{fig:results}
\end{figure}
\vspace*{-0.2cm}

The negative dataset $S_-$ necessary for RP and EA is composed of $\Sigma^1 \cup \Sigma^2$ with $\Sigma = \{a,b,c\}$. CC and TR do not require any negative set.
We evaluated our methods on each subset of $k$ distinct languages for $k$ from 2 to 5.  
The number of input languages $k$ has been given as parameter to the learning algorithms, to return $k$ output languages.
We show in Figure \ref{fig:results}a to \ref{fig:results}d the purity as a function of the density, depending on the number of languages to learn $k$.

The method CC performs remarkably well even with small training sets, whereas TR fails to generalize repeated patterns only having a cardinality of at least 2. For RP, we see that a dataset size of at least 5 strings improves the results. Concerning EA, we see that it underperforms whenever the number of DFAs to learn exceeds 4.
Surprisingly, it seems that approaches which learn from positive examples achieve better results than those needing positive and negative examples, especially when the number of DFAs to learn is too large.

\vspace*{-0.2cm}
\section{Discussion}
\label{sec:discussion}
\vspace*{-0.2cm}

We modified state-of-the-art algorithms and applied it to a set of \textit{simple} languages. We wonder on what class(es) of language(s) our methods work the best, and also would like to explore some properties of the learning algorithms.

\paragraph{Class of languages}

It is easy to see that CC, TR and RP will fail to distinguish $\{w \in \{a\}^*\ |\ |w| \mod 2 = 0\}$ from $\{w \in \{a\}^*\ |\ |w| \mod 2 = 1\}$ when they both are in $S_+$. Indeed, by construction of these algorithms, nothing ensures that odd or even repetitions of a pattern will be differentiated. However, they will properly recognize regular languages for which there is an integer $n$ such that for all words $x, y, z$ and integers $m \geq n$ then $xy^mz \in L$ iff $xy^nz \in L$. In other words, star-free languages can be easily recognized by our methods. However, EA offers more liberty and can possibly recognize some of the non-star-free languages (such as odd/even number of a's) since one can imagine a small dataset and a high number of clusters to find. The identification of different numbers of repetitions does not seem impossible.

\paragraph{Properties of RPNI-Splitting} We reflect on 2 properties of RPNI-splitting.

\begin{example}
Let us consider $\mathcal{A}$ as the canonical automaton returned by standard RPNI with $S_+$ and $S_-$ as learning sample. Let us consider the set of strings $s_+ \subset S_+$. Let $\mathcal{B}$ the automaton returned by standard RPNI with $s_+$ and $S_-$ as learning sample. Then
\[
s_+ \subset S_+ \not\implies L(\mathcal{B}) \subset L(\mathcal{A})
\]
\end{example}
Two explanations :
\begin{enumerate}
  \item By example in Appendix \ref{app:ex-rpni}. We gather 3 DFAs (Figure \ref{fig:threeoutputDFAs}b) by \texttt{RPNI-splitting}, and 1 DFA (Figure \ref{fig:threeoutputDFAs}a) by \texttt{standardRPNI}. Let us denote $\mathcal{A}$ the DFA in Figure \ref{fig:threeoutputDFAs}a and $\mathcal{B}$ the third DFA in Figure \ref{fig:threeoutputDFAs}b. $L(\mathcal{A})$ is gathered from $S_+ = \{aa,aaaa,aaaaaa,aaaaaaa,abba,$ $abbba,abbbba,abab,ababab,abababab\}$ and $L(\mathcal{B})$ is gathered from $s_+ = \{abab,ababab,$ $ abababab\}$. We see that $\{\lambda,bba\} \in L_1$ whereas $\{\lambda,bba\} \notin L_0$.
  \item During state merging between a blue state $b$ and a red state $r$, we know that a sub-DFA can be derived from the set of all reachable accepting states from $b$ as positive sample, and $S_-$ as negative sample. This set of states is a subset of $S_+$ and does not offer any guarantees about the language inclusion of the learned automata. Indeed, one could do more generalization than the other, depending on the sampling.
\end{enumerate}

\begin{example}
Let us consider $\mathcal{O}$ as the set of automata returned by Algorithm \ref{algo:rpnisplit} (RPNI-splitting). Let us consider $\mathcal{A}$ as the canonical automaton returned by standard RPNI. Finally, let $L(\mathcal{X})$ be the language accepted by automaton $\mathcal{X}$. Then 
\[
L(\mathcal{A}) \neq \bigcup_{O_i \in \mathcal{O}} L(O_i)
\]
\end{example}
By construction, RPNI-splitting extracts sub-solutions such that each sub-DFA is computed from a subset of the positive sample. A first intuition would be to say that the union of all sub-DFAs would be equal to the unique automaton returned by standard RPNI. Nevertheless, each sub-solution undergoes its own merging operations, that can lead to more or less generalizations in comparison to the unique solution.

\paragraph{Bound on the number of DFAs with RPNI-Splitting}
By construction, RPNI-splitting cannot render more DFAs than the number of state merging operations standard RPNI would perform. One could easily imagine that even with a large $k$ splitting factor, the number of returned DFAs will be bounded.

\section{Conclusion}
\label{sec:conclusion}
In this paper, we describe two innovative methods to learn several languages from labeled strings.
The first one consists in performing PTA-splitting, that is to say, a heuristic-based derivation of sub-solutions from the original state merging algorithms.
The later consists in an evolutionary algorithm which fitness function combines the accuracy of the DFA and the number of sub-DFAs that can be derived from it, by parsing the positive strings in the DFA and counting how many different transition sequences exist. 
From a practical perspective, we see that the gathering of more than one automaton can be meaningful since a too generalized solution is not always the best. This is notably the case in some industrial settings.
We also evaluated these methods as well as the methods described in \cite{learnaut}, and conclude that the different methods provide comparable results when the number of DFAs to learn is small. However, in case the number of DFAs to learn is great, the inference of multiple languages is of better quality from positive examples only.  We will then investigate to what extent compression can be useful for grammatical inference.

As further works, we will investigate the definition of a metric that would compare the different DFAs provided by the different methods. We highlight the necessity of such a measure since its absence prevents us from a proper evaluation of our methods. We think that the sole use of the predictive factor of the different automata is not a good enough measure. The finding of relevant heuristics or statistics to that end promises to be complicated since it is difficult to determine whether a solution is better than another because it generalizes too much/not enough, gives too many/too few sub-DFAs, etc. We probably face here a philosophical question from a theoretical perspective, but depending on the applications, expert knowledge can decide which heuristic is relevant. We would also like to adapt automata learning as an SMT problem \citep{smetsers2017grammatical} to the learning of several languages. Finally, we think that we can develop active learning techniques in order to learn several systems by querying a teacher and learning more than one System Under Test (SUT). Of course, this would imply to redefine Minimally Adequate Teachers (MAT), and also the different queries that the learner could send to the teacher.

\section*{Acknowledgments}
This research is supported by the Dutch Technology Foundation STW under the Robust CPS program (project 12693).

\newpage
\appendix

\section{Example for RPNI-splitting.}
\label{app:ex-rpni}
\vspace*{-0.2cm}
Let us consider a positive sample $S_+=\{aaaa,aaaaaa,aaaaaaa,abba,abbba,abbbba,abab,$ $ababab,abababab\}$ and a negative sample $S_-=\{a,b,bb\}$. We define a value $k = 5$. During the first consequent merge operation, states $aa$ and $aaa$ are merged.
From the resulting DFA, we extract 3 accepting states \{aaaa,aaaaaa,aaaaaaa\} from which we build a sub-solution with \texttt{standardRPNI}.
Then, \texttt{RPNI-splitting} is run with $S_+=\{abba,abbba,abbbba,$ $abab,ababab,abababab\}$ and $S_-=\{a,b,bb,aaaa,$ $aaaaaa,aaaaaaa\}$. A consequent merge operation happens when states $ab$ and $abb$ are merged. 
From the resulting DFA, we extract 3 accepting states \{abba,abbba,abbbba\} from which we build a sub-solution with \texttt{standardRPNI}.
Then, \texttt{RPNI-splitting} is run with $S_+=\{abab,ababab,abababab\}$ and $S_-=\{a,b,bb,aaaa,aaaaaa,aaaaaaa,abba,$ $abbba,abbbba,\}$. No consequent merge is detected, so the result of \texttt{standardRPNI} is computed on this sample.
We gather at the termination of the algorithm 3 different automata. 

\vspace*{-0.9cm}
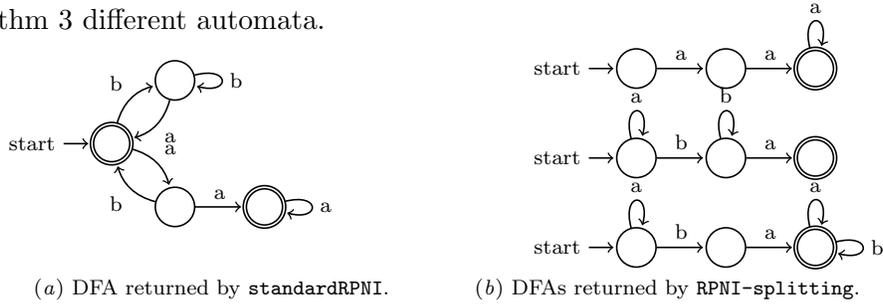
\begin{figure}[h!]
\begin{scriptsize}
\centering     
\subfigure[DFA returned by \texttt{standardRPNI}.]{
    \begin{tikzpicture}[->,shorten >=1pt,auto,node distance=1.7cm,semithick]
      \node[initial,accepting,state,scale=0.7] (init)                    {};
      \node[state,scale=0.7]          			(A) [below right of=init] {};
      \node[state,scale=0.7]          (B) [above right of=init] {};
      \node[state,accepting,scale=0.7]          (AA) [right of=A] {};
      \path (init) edge  [bend left]            node {b} (B)
            (B)    edge	[loop right] node {b} (B)
            (B)    edge	[bend left]			 node {a} (init)
            (init) edge  [bend left]            node {a} (A)
              (A) edge  [bend left]            node {b} (init)
              (A) edge  []            node {a} (AA)
              (AA) edge  [loop right]            node {a} (AA)
              (A)   edge [loop below,draw=none] node {} (A)
            ;
    \end{tikzpicture}
    \hspace*{1cm}
}
\subfigure[DFAs returned by \texttt{RPNI-splitting}.]{
\hspace*{1cm}
\begin{tikzpicture}[->,shorten >=1pt,auto,node distance=1.7cm,semithick]
      \node[initial,state,scale=0.7] (init)                    {};
      \node[state,scale=0.7]          (A) [right of=init] {};
      \node[state,accepting,scale=0.7]          (AA) [right of=A] {};
      	 \node[initial,state,scale=0.7] (init2)  [below of=init]                  {};
      \node[state,scale=0.7]          (B2) [right of=init2] {};
       \node[state,accepting,scale=0.7]         (BA2) [right of=B2] {};
        \node[initial,state,scale=0.7] (init3)    [below of=init2]                 {};
      \node[state,scale=0.7]          (B3) [right of=init3] {};
       \node[state,accepting,scale=0.7]         (BA3) [right of=B3] {};      
      \path (init) edge              node {a} (A)
            (A)    edge				 node {a} (AA)
            (AA)   edge [loop above] node {a} (AA)
            (init2) edge [loop above] node {a} (init2)
      		(init2) edge [] node {b} (B2)
            (B2)    edge	[loop above] node {b} (B2)       
            (B2)    edge	[]	node {a} (BA2)
            (init3) edge [loop above] node {a} (init3)
      		(init3) edge [] node {b} (B3)
            (B3)    edge	[]	node {a} (BA3)
            (BA3)    edge	[loop above]	node {a} (BA3)
            (BA3)    edge	[loop right]	node {b} (BA3)
            ;
    \end{tikzpicture}
}
\vspace*{-0.2cm}
\caption{Result of \texttt{standardRPNI} compared to \texttt{RPNI-splitting}.}
\label{fig:threeoutputDFAs}
\end{scriptsize}
\end{figure}

\vspace*{-0.5cm}
\section{Example for transitions clustering.}
\label{app:transition-clust}
\vspace*{-0.2cm}

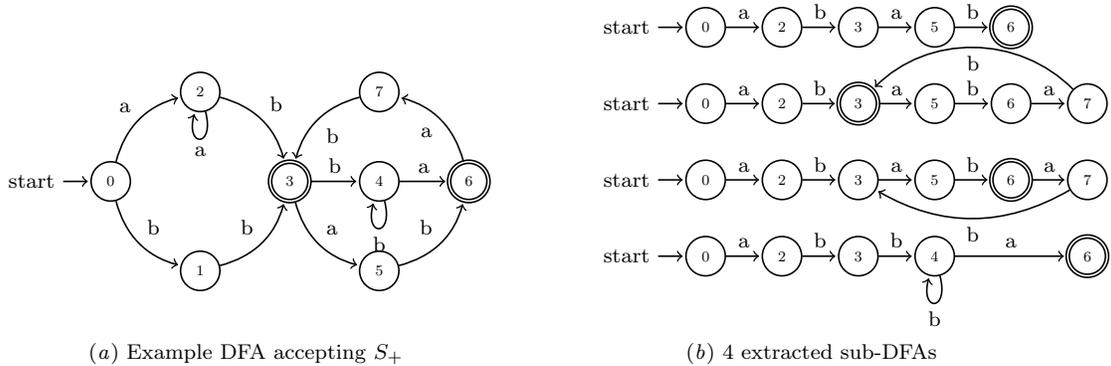
\begin{figure}[h!]
\begin{scriptsize}
\centering     
\subfigure[Example DFA accepting $S_+$]{
\begin{tikzpicture}[->,shorten >=1pt,auto,node distance=1.7cm,semithick]
      \node[initial,state,scale=0.7] 				(0)                    {0};
      \node[state,scale=0.7,draw=none] 				(fake) [right of=0]    {};
      \node[state,scale=0.7]          				(1) [below of=fake] {1};
      \node[state,scale=0.7]          				(2) [above of=fake] {2};
      \node[state,accepting,scale=0.7]         		(3) [right of=fake] {3};
      \node[state,scale=0.7]         				(4) [right of=3]	   {4};
      \node[state,scale=0.7]         				(5) [below of=4] 	   {5};
      \node[state,accepting,scale=0.7]  			(6) [right of=4]	   {6};
      \node[state,scale=0.7]  						(7) [above of=4]	   {7};
      \path (0) edge [bend right] node {b} (1)
      		(0) edge [bend left]  node {a} (2)
      		(1) edge [bend right] node {b} (3)
      		(2) edge [loop below] node {a} (2)
      		(2) edge [bend left]  node {b} (3)
      		(3) edge [] 		  node {b} (4)
      		(3) edge [bend right] node {a} (5)
      		(4) edge [loop below] node {b} (4)
      		(4) edge []			  node {a} (6)
      		(5) edge [bend right] node {b} (6)
      		(6) edge [bend right] 		  node {a} (7)
      		(7) edge [bend right] node {b} (3)
            (1)   edge [loop below,draw=none] node {} (1)
            ;
\end{tikzpicture}
}
\subfigure[4 extracted sub-DFAs]{
	\hspace*{1cm}
		\begin{tikzpicture}[->,shorten >=1pt,auto,node distance=1.45cm,semithick]
      \node[initial,state,scale=0.7] 				(0a)                    {0};
      \node[state,scale=0.7,color=black]          				(2a) [right of=0a] {2};
      \node[state,scale=0.7,color=black]         		(3a) [right of=2a] {3};
      \node[state,scale=0.7,color=black]         				(5a) [right of=3a] 	   {5};
      \node[state,accepting,scale=0.7,color=black]  			(6a) [right of=5a]	   {6};
      \node[initial,state,scale=0.7] 				(0b)     [below of=0a]               {0};
      \node[state,scale=0.7,color=black]          				(2b) [right of=0b] {2};
      \node[state,accepting,scale=0.7,color=black]         		(3b) [right of=2b] {3};
      \node[state,scale=0.7,color=black]         				(5b) [right of=3b] 	   {5};
      \node[state,scale=0.7,color=black]  			(6b) [right of=5b]	   {6};
      \node[state,scale=0.7,color=black]  			(7b) [right of=6b]	   {7};
       \node[initial,state,scale=0.7] 				(0c)     [below of=0b]               {0};
      \node[state,scale=0.7,color=black]          				(2c) [right of=0c] {2};
      \node[state,scale=0.7,color=black]         		(3c) [right of=2c] {3};
      \node[state,scale=0.7,color=black]         				(5c) [right of=3c] 	   {5};
      \node[state,accepting,scale=0.7,color=black]  			(6c) [right of=5c]	   {6};
      \node[state,scale=0.7,color=black]  			(7c) [right of=6c]	   {7};
       \node[initial,state,scale=0.7] 				(0d)      [below of=0c]              {0};
      \node[state,scale=0.7,color=black]          				(2d) [right of=0d] {2};
      \node[state,scale=0.7,color=black]         		(3d) [right of=2d] {3};
      \node[state,scale=0.7,color=black]         				(4d) [right of=3d] 	   {4};
      \node[state,accepting,scale=0.7,color=black]  			(6d) [below of=7c]	   {6};
      \path 
      		(0a) edge []  node {a} (2a)
      		(2a) edge []  node {b} (3a)
      		(3a) edge [] node {a} (5a)
      		(5a) edge [] node {b} (6a)
            (0b) edge []  node {a} (2b)
      		(2b) edge []  node {b} (3b)
      		(3b) edge [] node {a} (5b)
      		(5b) edge [] node {b} (6b)
      		(6b) edge [] node {a} (7b)
      		(7b) edge [bend right=45] node {b} (3b)
            (0c) edge []  node {a} (2c)
      		(2c) edge []  node {b} (3c)
      		(3c) edge [] node {a} (5c)
      		(5c) edge [] node {b} (6c)
      		(6c) edge [] node {a} (7c)
      		(7c) edge [bend left] node {b} (3c)
            (0d) edge []  node {a} (2d)
      		(2d) edge []  node {b} (3d)
      		(3d) edge [] node {b} (4d)
      		(4d) edge [loop below] node {b} (4d)
      		(4d) edge [] node {a} (6d)
            ;
	\end{tikzpicture}
}
\vspace*{-0.2cm}
\caption{Sub-DFAs extraction by transitions clustering.}
\label{fig:4outputDFAs}
\end{scriptsize}
\end{figure}

Given the DFA in Figure \ref{fig:4outputDFAs}a and a positive dataset $S_+ = \{abab,ababab,abababab,abbba,$ $ abbbbba,abbbbbbba\}$, let us compute how many sub-DFAs we can extract, computing the different transitions taken by the strings in $S_+$.
String $abab$ is accepted using transitions (0,2), (2,3), (3,5) and (5,6) with 6 being an accepting state.
String $ababab$ is accepted using transitions (0,2), (2,3), (3,5), (5,6) (6,7) and (7,3) with 3 being an accepting state.
String $abababab$ is accepted using transitions (0,2), (2,3), (3,5), (5,6) (6,7) and (7,3) with 6 being an accepting state.
Strings $abbba,abbbbba$ and $abbbbbbba$ are accepted using transitions (0,2), (2,3), (3,4), (4,4) and (4,6) with 6 being an accepting state.
We gather here 4 different DFAs.

\newpage
\bibliography{biblio}

\begin{thebibliography}{13}
\providecommand{\natexlab}[1]{#1}
\providecommand{\url}[1]{\texttt{#1}}
\expandafter\ifx\csname urlstyle\endcsname\relax
  \providecommand{\doi}[1]{doi: #1}\else
  \providecommand{\doi}{doi: \begingroup \urlstyle{rm}\Url}\fi

\bibitem[Carrasco and Oncina(1994)]{carrasco1994learning}
Rafael~C Carrasco and Jos{\'e} Oncina.
\newblock Learning stochastic regular grammars by means of a state merging
  method.
\newblock In \emph{International Colloquium on Grammatical Inference}, pages
  139--152. Springer, 1994.

\bibitem[de~la Higuera(2010)]{cdlh}
Colin de~la Higuera.
\newblock \emph{Grammatical inference: learning automata and grammars}.
\newblock Cambridge University Press, 2010.

\bibitem[Deb et~al.(2002)Deb, Agrawal, Pratap, and Meyarivan]{nsga2}
Kalyanmoy Deb, Samir Agrawal, Amrit Pratap, and T.~Meyarivan.
\newblock A fast and elitist multiobjective genetic algorithm: Nsga-ii.
\newblock \emph{IEEE Transactions on Evolutionary Computation}, 6\penalty0
  (2):\penalty0 182--197, Apr 2002.

\bibitem[Dupont(1994)]{dupont94}
Pierre Dupont.
\newblock Regular grammatical inference from positive and negative samples by
  genetic search: the gig method.
\newblock In \emph{Grammatical Inference and Applications}, pages 236--245.
  Springer Berlin Heidelberg, 1994.

\bibitem[Heule and Verwer(2010)]{heule2010exact}
Marijn~JH Heule and Sicco Verwer.
\newblock Exact dfa identification using sat solvers.
\newblock In \emph{Proceedings of the 10th International Colloquium on
  Grammatical Inference}, pages 66--79, 2010.

\bibitem[Lambeau et~al.(2008)Lambeau, Damas, and Dupont]{lambeau2008state}
Bernard Lambeau, Christophe Damas, and Pierre Dupont.
\newblock State-merging dfa induction algorithms with mandatory merge
  constraints.
\newblock \emph{Grammatical Inference: Algorithms and Applications}, pages
  139--153, 2008.

\bibitem[Lang et~al.(1998)Lang, Pearlmutter, and Price]{lang1998results}
Kevin~J Lang, Barak~A Pearlmutter, and Rodney~A Price.
\newblock Results of the abbadingo one dfa learning competition and a new
  evidence-driven state merging algorithm.
\newblock In \emph{International Colloquium on Grammatical Inference}, pages
  1--12. Springer, 1998.

\bibitem[{Linard} et~al.(2017){Linard}, {Smetsers}, {Vaandrager}, {Waqas}, {van
  Pinxten}, and {Verwer}]{learnaut}
Alexis {Linard}, Rick {Smetsers}, Frits {Vaandrager}, Umar {Waqas}, Joost {van
  Pinxten}, and Sicco {Verwer}.
\newblock {Learning Pairwise Disjoint Simple Languages from Positive Examples}.
\newblock \emph{ArXiv e-prints arXiv:1706.01663}, June 2017.

\bibitem[Neider(2012)]{Neider2012}
Daniel Neider.
\newblock Computing minimal separating dfas and regular invariants using sat
  and smt solvers.
\newblock In Supratik Chakraborty and Madhavan Mukund, editors, \emph{Automated
  Technology for Verification and Analysis}, pages 354--369. Springer Berlin
  Heidelberg, 2012.

\bibitem[Oncina and Garcia(1992)]{oncina1992identifying}
Jos{\'e} Oncina and Pedro Garcia.
\newblock Identifying regular languages in polynomial time.
\newblock \emph{Advances in Structural and Syntactic Pattern Recognition},
  5\penalty0 (99-108):\penalty0 15--20, 1992.

\bibitem[Smetsers(2017)]{smetsers2017grammatical}
Rick Smetsers.
\newblock Grammatical inference as a satisfiability modulo theories problem.
\newblock \emph{arXiv preprint arXiv:1705.10639}, 2017.

\bibitem[Wever et~al.(2017)Wever, van Rooijen, and Hamann]{rooijen}
Marcel Wever, Lorijn van Rooijen, and Heiko Hamann.
\newblock Active coevolutionary learning of requirements specifications from
  examples.
\newblock In \emph{Proceedings of the Genetic and Evolutionary Computation
  Conference}, GECCO '17, pages 1327--1334. ACM, 2017.

\bibitem[Zhang et~al.(2017)Zhang, Lin, Wang, and Verwer]{ZHANG20172353}
Yihuan Zhang, Qin Lin, Jun Wang, and Sicco Verwer.
\newblock Car-following behavior model learning using timed automata.
\newblock \emph{IFAC-PapersOnLine}, 50\penalty0 (1):\penalty0 2353 -- 2358,
  2017.
\newblock 20th IFAC World Congress.

\end{thebibliography}

\end{document}